\documentclass[conference]{IEEEtran}
\IEEEoverridecommandlockouts

\usepackage{graphicx}
\usepackage{subcaption}
\usepackage{amsmath}
\usepackage{amssymb}

\usepackage{color}

\DeclareMathOperator*{\diag}{diag}

\usepackage[makeroom]{cancel}

\usepackage{algorithm}
\usepackage{algpseudocode}

\usepackage[noadjust]{cite}

\usepackage{mathrsfs}

\usepackage{soul}

\usepackage{bm}
\begin{document}

\title{Oscillation energy based sensitivity analysis and control for multi-mode oscillation systems
\thanks{This material is based on work supported by the National Science Foundation under Grant No. 1509114. This work is also supported by the Engineering Research Center Program of the National Science Foundation and the Department of Energy under NSF Award No. EEC-1041877 and the CURENT Industry Partnership Program.}
}

\author{\IEEEauthorblockN{Horacio Silva-Saravia, Yajun Wang, H\'{e}ctor Pulgar-Painemal, Kevin Tomsovic}
\IEEEauthorblockA{Department of Electrical Engineering and Computer Science\\
University of Tennessee, Knoxville, TN, 37996\\
Email: hsilvasa@vols.utk.edu, ywang139@vols.utk.edu, hpulgar@utk.edu, ktomsovi@utk.edu}}

\maketitle
 
\begin{abstract}
This paper describes a novel approach to analyze and control systems with multi-mode oscillation problems. Traditional single dominant mode analysis fails to provide effective control actions when several modes have similar low damping ratios. This work addresses this problem by considering all modes in the formulation of the system kinetic oscillation energy. The integral of energy over time defines the total action as a measure of dynamic performance, and its sensitivity allows comparing the performance of different actuators/locations in the system to select the most effective one to damp the oscillation energy. Time domain simulations in the IEEE 9-bus system and IEEE 39-bus system verify the findings obtained by the oscillation energy based analysis. Applications of the proposed method in control and system planning are discussed. 
\end{abstract}


\begin{IEEEkeywords}
Damping control, eigenvalue sensitivity, inter-area oscillations, small-signal stability, energy storage, renewable energy, oscillation energy, action.
\end{IEEEkeywords}

\IEEEpeerreviewmaketitle

\section{Introduction}


  Electromechanical oscillations in power systems appear as result of energy exchanges between different groups of generators after a disturbance. These oscillations are unwanted because of the mechanical stress in generator shafts, power congestion in the transmission system and the potential risk of instability. Traditionally, control efforts are done to guarantee higher damping ratios for these oscillations, with special consideration for inter-area oscillations, which involve a larger part of the system. Although for the current scenario these control actions fulfill the system requirements, research on control schemes and a better understanding of the oscillation problem need to continue as higher penetration of renewable energy (RE)---and decommission of traditional generation---will take place, making the oscillation problem more challenging.

Selected power system stabilizers (PSSs) and feedback signals using residue analysis and mode controllability/observability have been traditionally employed to damp electromechanical oscillations \cite{Martins1990}, \cite{Wang1999}. Residue analysis focuses on selecting the input/output pairs that are most sensitives to displace a targeted eigenvalue to the left side of the complex plane. Similar approaches have been extended for the location of flexible AC transmission systems (FACTS) and energy storage systems (ESS) \cite{Silva2016}. Special efforts have also been able to estimate the system inertia distribution \cite{Yajun2017} and find relationships with the location of electrically-interfaced resources (EIRs) to damp inter-area oscillations \cite{Pulgar2017}. In these works, and the majority of real system studies \cite{Rimorov2016}, only single-mode analysis is performed. This is based on the assumption that there is only one dominant mode, which is not guaranteed in scenarios with high penetration of RE because of the reduction of the relative inertia in different areas of the system. To consider multi-mode analysis and achieve some arbitrary system performance, optimization-based techniques have been implemented to tune and design system controls \cite{Cai2005}. This has brought the idea of control allocation and coordination in power systems to distribute control effort among multiple actuators \cite{Ehsan2017}. However, these optimization-based techniques lack of physical interpretation and depend on arbitrary design parameters.

A different approach to study electromechanical oscillations consists of analyzing the kinetic oscillation energy of each machine by comparing the phase of selected energy modes to identify energy exchanges paths \cite{Jing1996}, \cite{Messina2001}. Recent efforts also consider the distribution of the kinetic energy, branch potential energy \cite{Yu2016} and the idea of energy dissipation and its relationship with oscillation damping \cite{Chen2014}. Although more meaningful in terms of the system physical interpretation, these works still fail to provide a performance index considering all system modes. Moreover, they do not provide direct comparison of different control actuators and the effects on the oscillation energy as a measure of system performance.

This paper proposes a new approach to study system oscillations---specially when considering high penetration of RE---by considering all oscillation modes in the formulation of the system kinetic oscillation energy. This formulation allows comparing the dynamic performance of control actuators/locations in the system by means of the total action sensitivities (TAS). The most sensitive actuators/locations are proven to provide the best dynamic response for the system. Simulations in the IEEE 9-bus system  and IEEE 39-bus systems verify the findings of this work. The paper is structured as follows. Section II describes the concepts of oscillation energy, action and total action sensitivity. Section III compares the traditional eigenvalue sensitivity analysis with the proposed method in the IEEE 9-bus test system and shows an application of the total action sensitivity in the IEEE 39-bus test system. Final remarks about applications of the analysis are also discussed. Finally, conclusions are presented in Section IV.

\section{Oscillation energy analysis}
\subsection{Oscillation energy and action}
Consider the linearized power system equations with $p$ synchronous generators and $n$ total number of states 

\begin{equation}
\Delta \dot{x} = A\Delta x
\label{eq:sol}
\end{equation}
By using the transformation $\Delta x=M \Delta z$, where $M=\{v_1,v_2,...v_n\}$ is the matrix of right eigenvectors, the system equations can be decoupled as: 
\begin{equation}
\Delta \dot{z} = \underbrace{M^{-1}AM}_{\Lambda}  \Delta z =\Lambda \Delta z
\label{eq:solz}
\end{equation}
Here $\Lambda=\diag \{\lambda_i\}$, where $\lambda_i$ is the i-th system eigenvalue. Thus,  the solution of each state of the decoupled system can be easily written in terms of its corresponding eigenvalue:

\begin{equation}
\Delta z = e^{\Lambda t} \Delta z_0 \rightarrow \Delta z_i = e^{\lambda_i t}\Delta z_{0i} \in \mathbb{C}, \forall i\in \{1,...,n\}
\label{eq:solzz}
\end{equation}
where $\Delta z_0=[\Delta z_{01},...,\Delta z_{0i},...,\Delta z_{on}]^T= M^{-1}\Delta x_0$. The kinetic energy of the linearized system becomes:

\begin{align}
E_k(t)=\sum_{j=1}^{p}  \frac{1}{2}J_j\Delta \omega^2_j  =&\frac{1}{2}\Delta x^T J \Delta x\\
  =&\frac{1}{2}(M\Delta z)^T J (M\Delta z)\\
  =&\frac{1}{2}\Delta z^T G\Delta z  \in \mathbb{R}
\label{eq:ke}
\end{align}
where the inertia matrix $J$ has nonzero elements only in the diagonal terms $J_{ii} ~ \forall~ i \in \Omega_\omega$, where $\Omega_\omega$ is the set of speed indices of all synchronous generators. The transformed inertia matrix $G=M^TJM$ is in general non diagonal and complex. Note that after a disturbance, the speed trajectories describe the oscillation energy  defined by equation \eqref{eq:ke} such that $E_k(t)>0 \quad\forall~t$ and $E_k$ is zero in steady state. Consider now the mathematical definition of action ($S$), which is typically represented by an integral over time and taken along the system trajectory \cite{Lanczos2012}. This integral has units of (energy)$\cdot$(time) and for our problem can be written as:


\begin{align}
S(\tau)=& \int_{0}^{\tau} E_k(t)dt=\int_{0}^{\tau} \frac{1}{2}(\Delta z^T G\Delta z)dt \in \mathbb{R}\\
   =&\int_{0}^{\tau} \frac{1}{2}(e^{\Lambda t} \Delta z_0)^T G(e^{\Lambda t} \Delta z_0)dt\\
 =&\frac{1}{2}\int_0^\tau \left(\sum_{j=1}^n\sum_{\substack{i=1}}^n e^{(\lambda_i+\lambda_j)t} z_{0i}z_{0j}g_{ij} \right) dt
\label{eq:D}
\end{align}
%

%
where $z_{0i}$ is the i-th element of $\Delta z_0$ and $g_{ij}$ is the entry in the i-th row and j-th column of $G$. 
The action evaluated at a fixed time $\tau$ becomes: 

\begin{align}
S(\tau) =&\frac{1}{2}\sum_{j=1}^n\sum_{i=1}^n\frac {e^{(\lambda_i+\lambda_j)t}}{(\lambda_i+\lambda_j)} z_{0i}z_{0j}g_{ij}\bigg\rvert_{0}^{\tau}
\label{eq:Dsol}
\end{align}
Considering stable modes, the total action until the oscillations vanish is obtained as,

\begin{align}
S_{\infty} =&\lim_{\tau \to \infty} S(\tau) = -\frac{1}{2}\sum_{j=1}^n\sum_{i=1}^n\frac {z_{0i}z_{0j}g_{ij}}{(\lambda_i+\lambda_j)} 
\label{eq:DsolT}
\end{align}

\subsection{Total action sensitivity (TAS)}
Assume that a damping control device is virtually installed in the system. The dynamics of this controller are fast and can be represented as a proportional gain $\theta_k$. Consider the analysis of the effect of the control gain $\theta_k$ in the total action, which is a measure of how quick the oscillation energy is damped. The sensitivity of the total action with respect to the control gain is expressed as:

\begin{align}
\frac{\partial S_{\infty}}{\partial \theta_k} =&-\sum_{j=1}^n\sum_{i=1}^n\frac {z_{0j}g_{ij}}{(\lambda_i+\lambda_j)}\frac{\partial z_{0i}}{\partial \theta_k}-\sum_{j=1}^n\sum_{i=1}^n\frac {z_{0i}z_{0j}}{2(\lambda_i+\lambda_j)}\frac{\partial g_{ij}}{\partial \theta_k} \nonumber \\
  &+\sum_{j=1}^n\sum_{i=1}^n\frac {z_{0i}z_{0j}g_{ij}}{2(\lambda_i+\lambda_j)^2}(\frac{\partial \lambda_i}{\partial \theta_k}+\frac{\partial \lambda_j}{\partial \theta_k})
\label{eq:sens}
\end{align}
where $\partial z_{oi}/\partial \theta_k$ and $\partial g_{ij}/\partial \theta_k$ are the entries of the following vector and matrix, respectively:
\begin{align}
\frac{\partial z_o}{\partial \theta_k} =& \frac{\partial M^{-1}}{\partial \theta_k} x_0\\
\frac{\partial G}{\partial \theta_k} =& \frac{\partial M^T}{\partial \theta_k} JM+M^T J\frac{\partial M}{\partial \theta_k}
\end{align}

Calculations of the eigenvector derivatives are obtained by solving a set of linear equations that are a function of the eigenvalues, their derivatives, the eigenvectors and the system matrix derivative \cite{Friswell1995}. Similarly, eigenvalue sensitivities can be calculated by means of the residue or equivalently using the concepts of mode controllability and mode observability \cite{Silva2016}, \cite{Pulgar2017}. 
For simplicity, equation \eqref{eq:sens} can be rearranged as a linear combination of the eigenvalue sensitivities plus one term that depends on the eigenvector sensitivities.

 \begin{equation}
\frac{\partial S_{\infty}}{\partial \theta_k}=\alpha_k+\sum_{i=1}^n \beta_i \frac{\partial \lambda_i}{\partial \theta_k}
\label{eq:dis}
 \end{equation}
where $\alpha_k \in \mathbb{R}$ is the summation of the first two terms in equation \eqref{eq:sens} and the modal coefficients $\beta_i$ are given by

\begin{equation}
\beta_i=\sum_{j=1}^n \frac{z_{0i}z_{0j}g_{ij}}{(\lambda_i+\lambda_j)^2}
\end{equation}

Note that $\partial S_{\infty}/\partial \theta_k$ is a real number, although $\beta_i$ and $\partial \lambda_i /\partial \theta_k$ are all complex quantities. Because of $E_k>0 \quad \forall ~t$, the best dynamic performance, from an energy point of view, occurs when $E_k$ quickly approaches to zero---which is equivalent to minimize the total action. Therefore, the control gain $\theta_k$ for which $\partial S_{\infty}/ \partial \theta_k<0$ and $|\partial S_{\infty}/ \partial \theta_k|$ is maximum, provides the optimal control solution.





 

\section{Simulation results and analysis}

The IEEE 9-bus system and IEEE 39-bus systems are used for simulations. Models and parameters are obtained from the library in DIgSILENT PowerFactory. A Battery Energy Storage System (BESS) is used to provide oscillation damping. Only a control gain is considered in the closed loop. The installation location of this BESS is analyzed for each bus $i$ at a time, and speed of the closest generator is used as feedback signal.

\subsection{Comparison between single mode analysis and oscillation energy analysis}

\begin{figure}[ht!]
\centering
\includegraphics[width=\columnwidth]{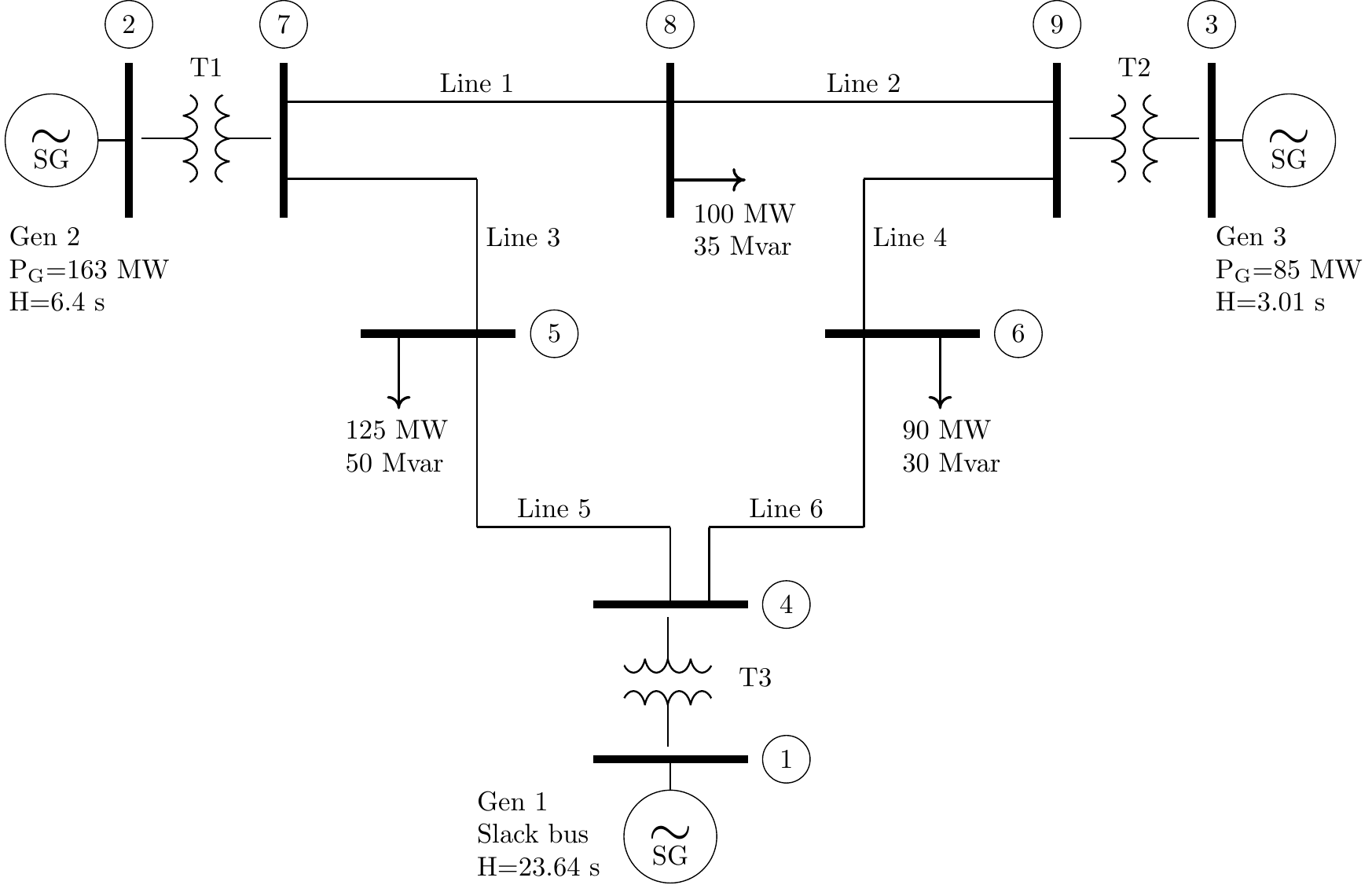}
\caption{3-machine, 9-bus system}
\label{fig:9bus}
\end{figure}

The IEEE 9-bus test system in Figure \ref{fig:9bus} is studied to show the advantage of the TAS over the traditional single-mode eigenvalue sensitivity analysis.


The system dynamics of the linearized model are dominated by two electromechanical modes: one local oscillation between Gen 2 and Gen 3 with an initial eigenvalue $\lambda_{23}=-0.027+j13.4$, and one inter-area oscillation between Gen 1 and (Gen 2, Gen 3) with an initial eigenvalue $\lambda_{123}=-0.038 + j8.73$. Note that both electromechanical modes have critical damping ratios of $0.19\%$ and $0.43\%$ respectively.

\subsubsection{Traditional eigenvalue sensitivity analysis}
\begin{figure}[ht!]
\centering
\includegraphics[width=0.9\columnwidth]{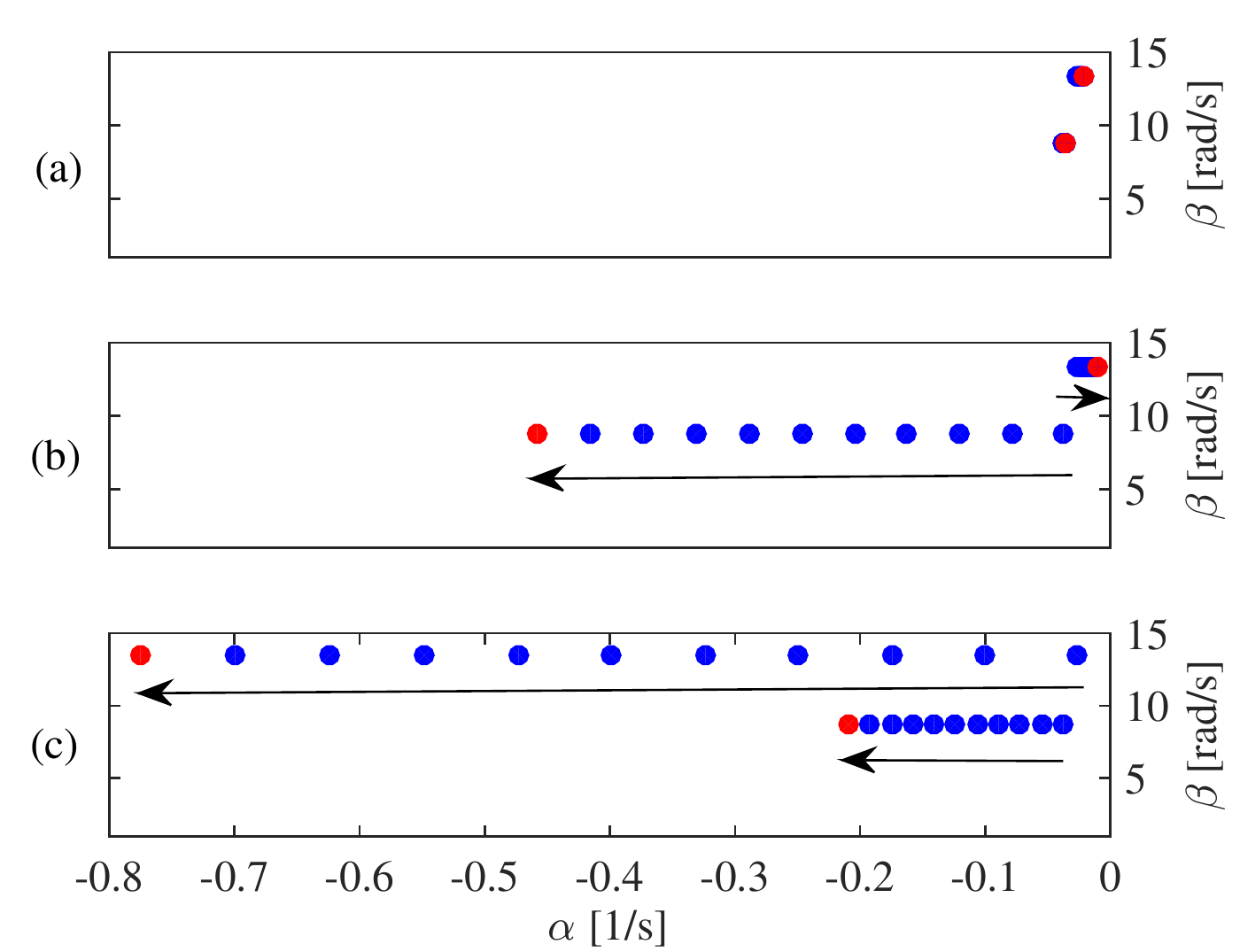}
\caption{Eigenvalue plot of the IEEE 9-bus system for prospective locations of BESS by increasing control gain. (a) Changing $\theta_4$, (b) changing $\theta_7$, (c) changing $\theta_9$}
\label{fig:eigenvalues}
\end{figure}

The location of a 100 BESS is studied at buses 4, 7 and 9 using the speed of generators 1, 2 and 3 as feedback signal, respectively. For each case, the control gain $\theta_i$ is increased from 0 to 50 in steps of 5, and the displacement of the local and inter-area mode are analyzed. 
Figure \ref{fig:eigenvalues} shows the eigenvalues displacement for an increasing control gain. A traditional approach would prioritize damping the inter-area oscillation, as both oscillations have low damping ratios and the inter-area oscillation involves more generators. Then, part (b) of the figure would be considered as the best case, i.e., increasing $\theta_7$ displaces further to the left-side plane the inter-area mode. The eigenvalue sensitivity shown in Table \ref{tab:sens} points out $\theta_7$ as the most effective gain to control the inter-area oscillation as well, while $\theta_9$ is more effective to control the local oscillation. Thus, from the point of view of eigenvalue sensitivity and each eigenvalue displacement, the prospective BESS location at bus 7 should be chosen to improve the system oscillations. However, as shown in the next subsection, this selection based on a single system eigenvalue is not always optimal.


\begin{table}[h!]
\renewcommand{\arraystretch}{1.3}
\caption{Eigenvalue sensitivities}
\label{tab:sens}
\centering
\begin{tabular}{lccc}
\hline
 & $|\partial \lambda_i/\partial \theta_4|$ & $|\partial \lambda_i/\partial \theta_7|$ & $|\partial \lambda_i/\partial \theta_9|$\\
\hline
$\lambda_{23}$ & $3.23\times 10^{-5}$ & $9.59\times10^{-5}$ & $\bm{0.0045}$ \\
$\lambda_{123}$ & $1.25\times 10^{-5}$ & $\bm{0.0025}$ & $0.0010$ \\
\hline
\end{tabular}
\end{table}

\subsubsection{Oscillation energy analysis}
The proposed oscillation energy and TAS analysis considering all modes is performed to provide insight about which BESS location---or combination of BESS locations---should be employed. Table \ref{tab:dis} shows the TAS for three different initial states disturbances $\Delta \omega_0 = (\Delta \omega_{01},\Delta \omega_{02},\Delta\omega_{03})$, where $\Delta\omega_{0j}$ denotes the initial speed deviation of machine $j$.

\begin{table}[h!]
\renewcommand{\arraystretch}{1.3}
\caption{Total action sensitivities}
\label{tab:dis}
\centering
\begin{tabular}{lccc}
\hline
 & $\partial S_{\infty}/\partial \theta_4$ & $\partial S_{\infty}/\partial \theta_7$ & $\partial S_{\infty}/\partial \theta_9$\\
\hline
$\Delta \omega_0^1=(0.01,0,-0.01)^T$ & $ -2.151 $ & $-21.42$ & $\bm{-40.18}$ \\
$\Delta \omega_0^2=(0,0.01,-0.01)^T$ & $-0.419$ & $-8.158 $ & $\bm{-57.56}$ \\
$\Delta \omega_0^3=(0.01,-0.01,0)^T$ & $-4.364$ & $\bm{-42.63}$ & $-17.11$ \\
\hline
\end{tabular}
\end{table}

\begin{figure}[ht!]
\centering
\includegraphics[width=0.9\columnwidth]{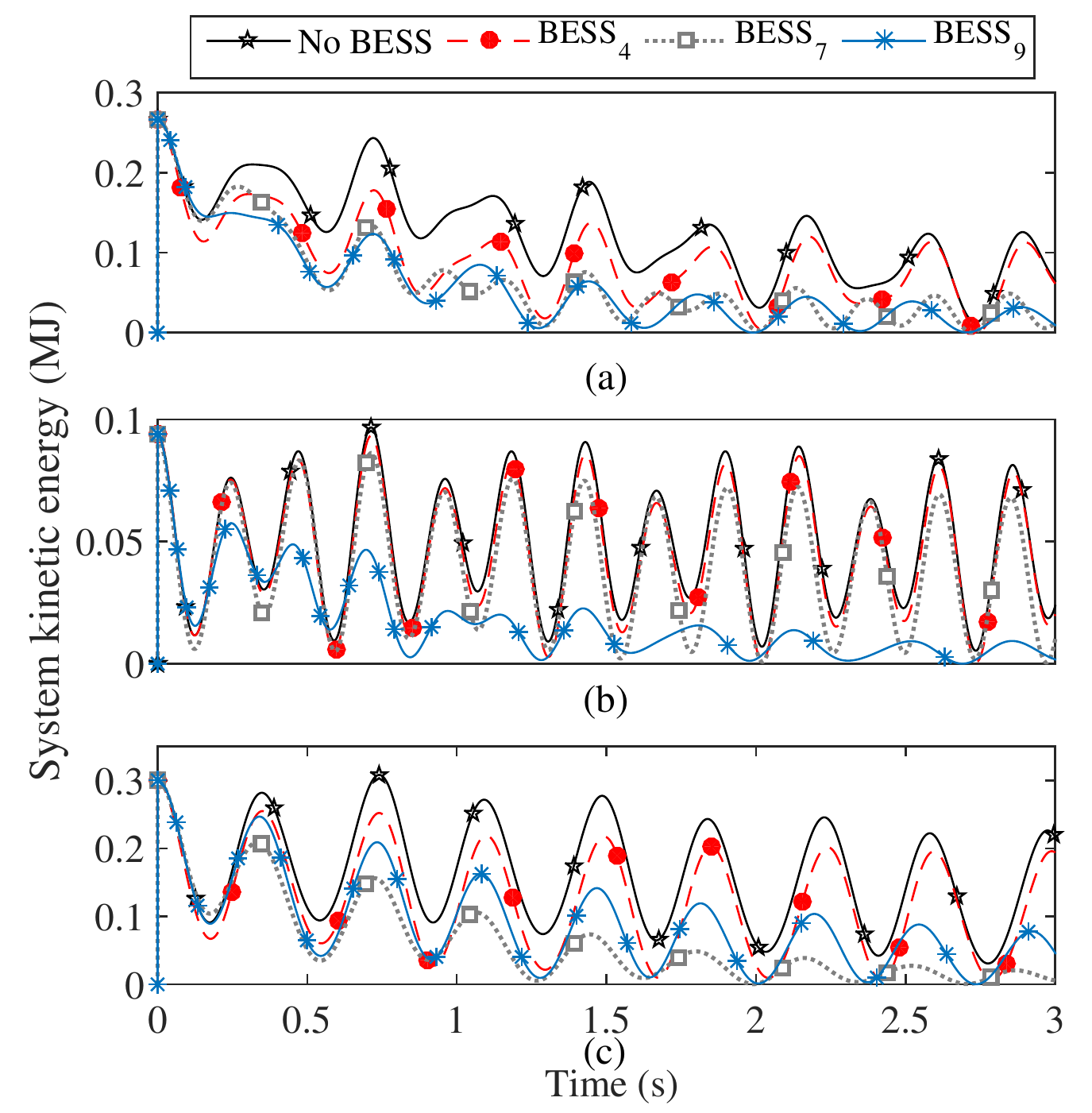}
\caption{System kinetic energy for different initial disturbances and BESS locations. (a) $\Delta \omega_0^1$, (b) $\Delta \omega_0^2$, (c) $\Delta \omega_0^3$  }
\label{fig:KE}
\end{figure}

 In order to verify the results obtained by the TAS analysis, time domain simulations are performed using the full set of nonlinear differential equations for each of the disturbances in Table \ref{tab:dis}. Figure \ref{fig:KE} shows the system kinetic energy for each prospective BESS location at a time.

The results from figure \ref{fig:KE} show agreement with those from Table \ref{tab:dis}. For the first and second disturbance in part (a) and (b) of the figure, the BESS located at bus 9 is more effective to damp the system kinetic energy while BESS located at bus 4 and 7 have marginal improvements. For the third disturbance shown in part (c) of the figure, the BESS located at bus 7 is the most effective to quickly drive the system to steady state. These differences occur because the disturbances excite modes in different proportions, aspect which is completely captured by the modal coefficients $\beta_i$ in equation \eqref{eq:dis}. To sum up, for some disturbances the single mode analysis fails to identify the best actuator/location, while the proposed approach is able to consider the combined effect of all eigenvalue displacements.


\subsection{Application: IEEE 39-bus test system}

\begin{figure}[ht!]
\centering
\includegraphics[scale=2.8]{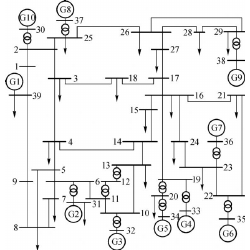}
\caption{IEEE 39-bus test system}
\label{fig:new_england}
\end{figure}

\begin{figure}[ht!]
\centering
\includegraphics[width=\columnwidth]{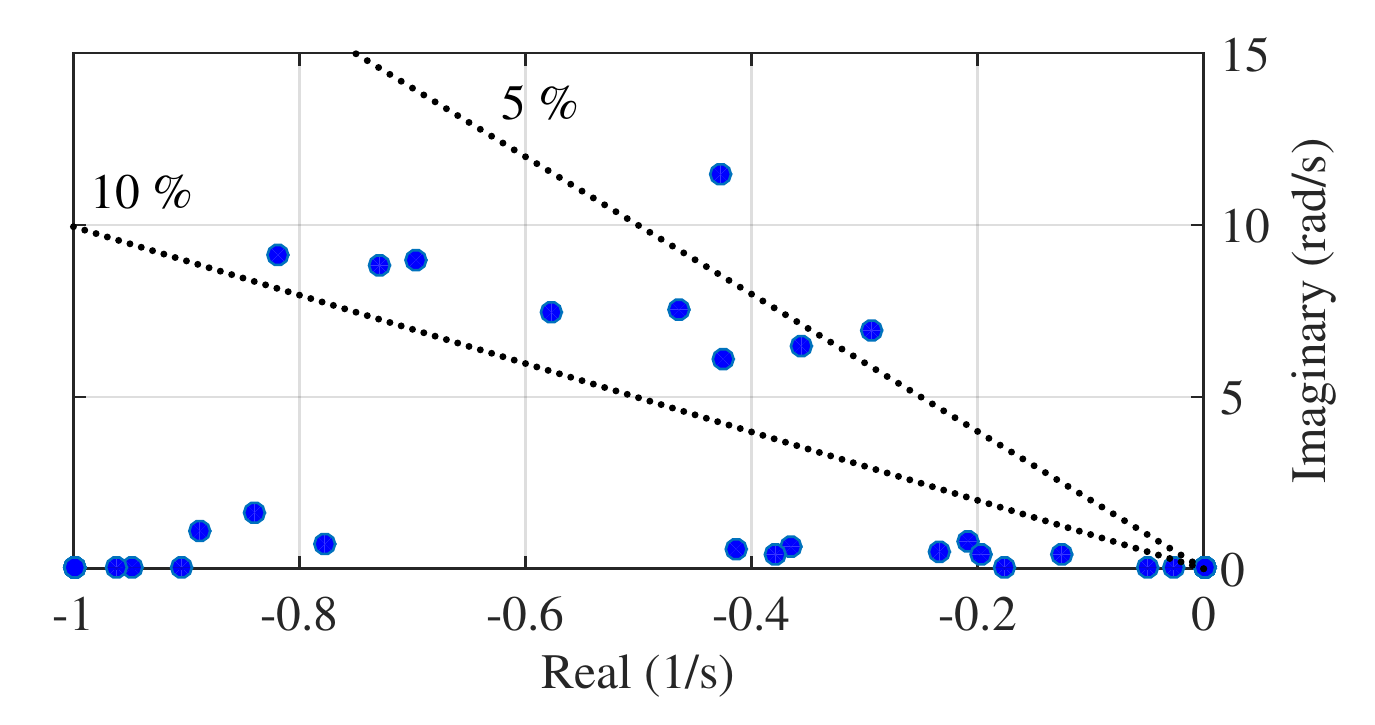}
\caption{System eigenvalues of the IEEE 39-bus test system}
\label{fig:evals}
\end{figure}

The TAS analysis is applied in the IEEE 39-bus test system shown in Figure \ref{fig:new_england}. The original inertia of generator $G_1$ is reduced to 30 s in a 100 MVA base to allow a more symmetric case. The dynamics of the system are described by the eigenvalues shown in Figure \ref{fig:evals}.  There are 9 electromechanical, most of them have damping ratios between 5\% and 10\% except one local mode of $G_1$ with frequency 11.5 rad/s and one inter-area mode between $G_{10}$ and ($G_2$, $G_3$, $G_9$) with frequency 6.9 rad/s.

Calculations for the TAS  analysis are performed for a 64 ms short-circuit at bus 12---fault clearing time for a two-cycle circuit breaker. Generator buses are chosen as prospective control buses. Machine speeds and angles are monitored and their values right after clearing the short-circuit are used as initial states in the sensitivities calculation. Table \ref{tab:dis39} shows the TAS $\partial S_{\infty}/ \partial \theta_k$ for each bus sorted from the best to the worst bus candidate to damp the oscillation energy. Additionally, the first and second column show the same calculation neglecting the sensitivity coefficient $\alpha$, i.e, assuming the eigenvector derivatives are zero, which comes from the assumption that the mode shapes are not affected by the control gain $\theta_k$.  As the table shows, both the exact and approximated results point out bus 39 as the best choice to control the system oscillations after this disturbance. Besides bus 30, all other buses play a similar role in damping the oscillation energy with relatively small differences. Note that the information provided in Table \ref{tab:dis39} can be also used to choose a set of optimal actuators in a centralized control scheme.


\begin{table}[h!]
\renewcommand{\arraystretch}{1.3}
\caption{Total action sensitivities for short-circuit at bus 12 in the IEEE 39-bus system}
\label{tab:dis39}
\centering
\begin{tabular}{cccc}
\hline
Bus & $\sum \beta_i \partial \lambda_i/\partial \theta_k$  & Bus & $\partial S_{\infty}/\partial \theta_k$\\
\hline
$39$ & $-0.0192$ & $39 $ &$-0.0136$ \\
$30$& $-0.0141$ & $30$ & $-0.0082$ \\
$36$ & $-0.0123$ & $32$  & $-0.0067$ \\
$35$ & $-0.0119$ & $31$  & $-0.0061$ \\
$34$ & $-0.0116$ & $37$  & $-0.0060$ \\
$38$ & $-0.0115$ & $36$  & $-0.0057$ \\
$33$ & $-0.0114$ & $38$  & $-0.0057$ \\
$37$ & $-0.0112$ & $33$  & $-0.0057$ \\
$32$ & $ -0.0111$ & $35$  & $-0.0056$ \\
$31$ & $-0.0104$ & $34$  & $-0.0056$ \\
\hline
\end{tabular}
\end{table}

\begin{figure}[ht!]
\centering
\includegraphics[width=\columnwidth]{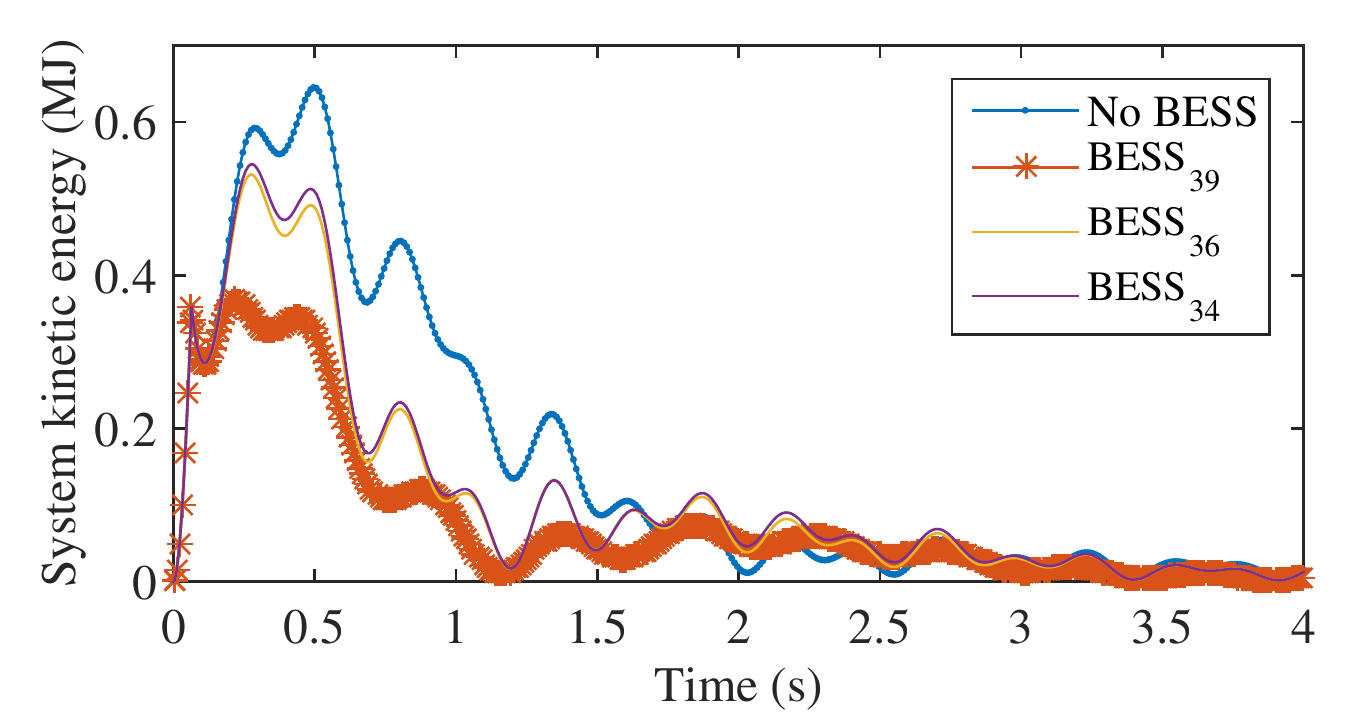}
\caption{System kinetic energy of the IEEE 39-bus system after a 64 ms short-circuit at bus 12}
\label{fig:time39}
\end{figure}

Time domain simulations are performed using the full set of nonlinear differential algebraic equations. A 200 MW BESS is connected at bus 39, 36 and 34 at a time to compare the results with those obtained by the TAS analysis. A delay block is added to the BESS control loop so it only reacts after the short circuit is cleared, which gives enough time to update the initial state vector in the TAS calculation and to send a signal to the best BESS location in the case of a centralized control scheme. Figure \ref{fig:time39} shows the system kinetic energy for the case without BESS and with BESS at each of the selected locations. The results show that the BESS at bus 39 is the most effective to quickly damp the oscillation energy, while the BESS at bus 36 and 34 have similar dynamic responses. Although this simulation is performed including the nonlinear equations and using a large droop gain for the BESS, still follows the expected results from the TAS analysis. Therefore, the TAS framework is proven to find the best actuator in the system. Note that, the accuracy of the results depend on the linearity of the eigenvalues and eigenvector trajectories. For nonlinear trajectories---usually for larger droop gains---second order sensitivities or linear piecewise approximation for the total action may be needed.

\subsection{Final remarks}

The proposed TAS analysis described in this paper successfully identifies the best control actuator/location in order to minimize the system kinetic energy variation over time. This can be used in several application, such as:
\begin{itemize}
\item Oscillation damping control allocation: on-line TAS evaluation can determine single or multiple actuators, either conventional or non conventional such as RE resources or energy storage.
\item Optimal tunning: phase lead compensator of different actuators can be tuned to optimize a total action-based cost function by changing the direction of eigenvalue trajectories.
\item System planning: off-line TAS analysis for common disturbances can lead to criteria  for the deployment of regulating devices to dynamically strength the system.
\end{itemize}

\section{Conclusion}

This paper describes an oscillation energy analysis to identify the best actuator/location in systems with multi-mode oscillation problems. By expressing the system kinetic energy in terms of the system eigenvalues and eigenvectors, and by calculating the sensitivity of the total action, an algebraic function of the initial states is obtained. The TAS results are validated in the IEEE 9-bus and IEEE 39-bus systems. Time domain simulations show that the TAS analysis is successful to provide the optimal solution in terms of the most effective actuator/location to quickly damp the oscillation energy, and therefore, damp all electromechanical oscillations. Promising applications of the TAS analysis in control and system planning are discussed.

\bibliographystyle{IEEEtran}
\bibliography{bare_conf}

\end{document}